\documentclass[conference, letterpaper]{IEEEtran}
\ifCLASSINFOpdf
\else
\fi
\usepackage[cmex10]{amsmath}
\ifCLASSINFOpdf
   \usepackage[pdftex]{graphicx}
\else
\fi

\usepackage[cmex10]{amsmath}
\usepackage{color}
\usepackage{fancyhdr}
\usepackage[caption=false,font=footnotesize]{subfig}

\renewcommand{\thispagestyle}[2]{}

\fancypagestyle{plain}{
        \fancyhead{}
        \fancyhead[C]{first page center header}
        \fancyfoot{}
        \fancyfoot[C]{first page center footer}
}
\usepackage{ctable,multirow,amsmath,graphicx}
\usepackage{url, comment}

\DeclareMathOperator*{\argmax}{arg\,max}

\begin{document}

%
\title{Neural Network Based Speaker Classification and Verification Systems with Enhanced Features\vspace{-.5cm}}

\author{\IEEEauthorblockN{Zhenhao Ge, Ananth N. Iyer, Srinath Cheluvaraja, Ram Sundaram, Aravind Ganapathiraju}
\IEEEauthorblockA{Interactive Intelligence Inc., Indianapolis, Indiana, USA\\
Email: \{roger.ge, ananth.iyer, srinath.cheluvaraja, ram.sundaram, aravind.ganapathiraju\}@inin.com}
}


\maketitle

\begin{abstract}
This work presents a novel framework based on feed-forward neural network for text-independent speaker classification and verification, two related systems of speaker recognition. With optimized features and model training, it achieves 100\% classification rate in classification and less than 6\% Equal Error Rate (ERR), using merely about 1 second and 5 seconds of data respectively. Features with stricter Voice Active Detection (VAD) than the regular one for speech recognition ensure extracting stronger voiced portion for speaker recognition, speaker-level mean and variance normalization helps to eliminate the discrepancy between samples from the same speaker. Both are proven to improve the system performance. In building the neural network speaker classifier, the network structure parameters are optimized with grid search and dynamically reduced regularization parameters are used to avoid training terminated in local minimum. It enables the training goes further with lower cost. In speaker verification, performance is improved with prediction score normalization, which rewards the speaker identity indices with distinct peaks and penalizes the weak ones with high scores but more competitors, and speaker-specific thresholding, which significantly reduces ERR in the ROC curve. TIMIT corpus with 8K sampling rate is used here. First 200 male speakers are used to train and test the classification performance. The testing files of them are used as in-domain registered speakers, while data from the remaining 126 male speakers are used as out-of-domain speakers, i.e. imposters in speaker verification.

\end{abstract}



%
\IEEEpeerreviewmaketitle

\section{Introduction}
\label{sec:intro}

Speaker recognition is a popular and broad topic in speech research over decades. It includes speaker detection, i.e. detecting if there is a speaker in the audio, speaker identification, i.e. identifying whose voice it is, speaker verification or authentication, i.e. verifying someone's voice. If the speaker set is closed, i.e. the audio must be from one of the enrolled speakers, then speaker identification is simplified to speaker classification. There are some other building blocks such speaker segmentation, clustering and diarization, which can be further developed based on the fundamental speaker recognition techniques.

\begin{figure}[htb]
 \centering
  \begin{tabular}{c}
 	\includegraphics[height=5cm, width=8cm]{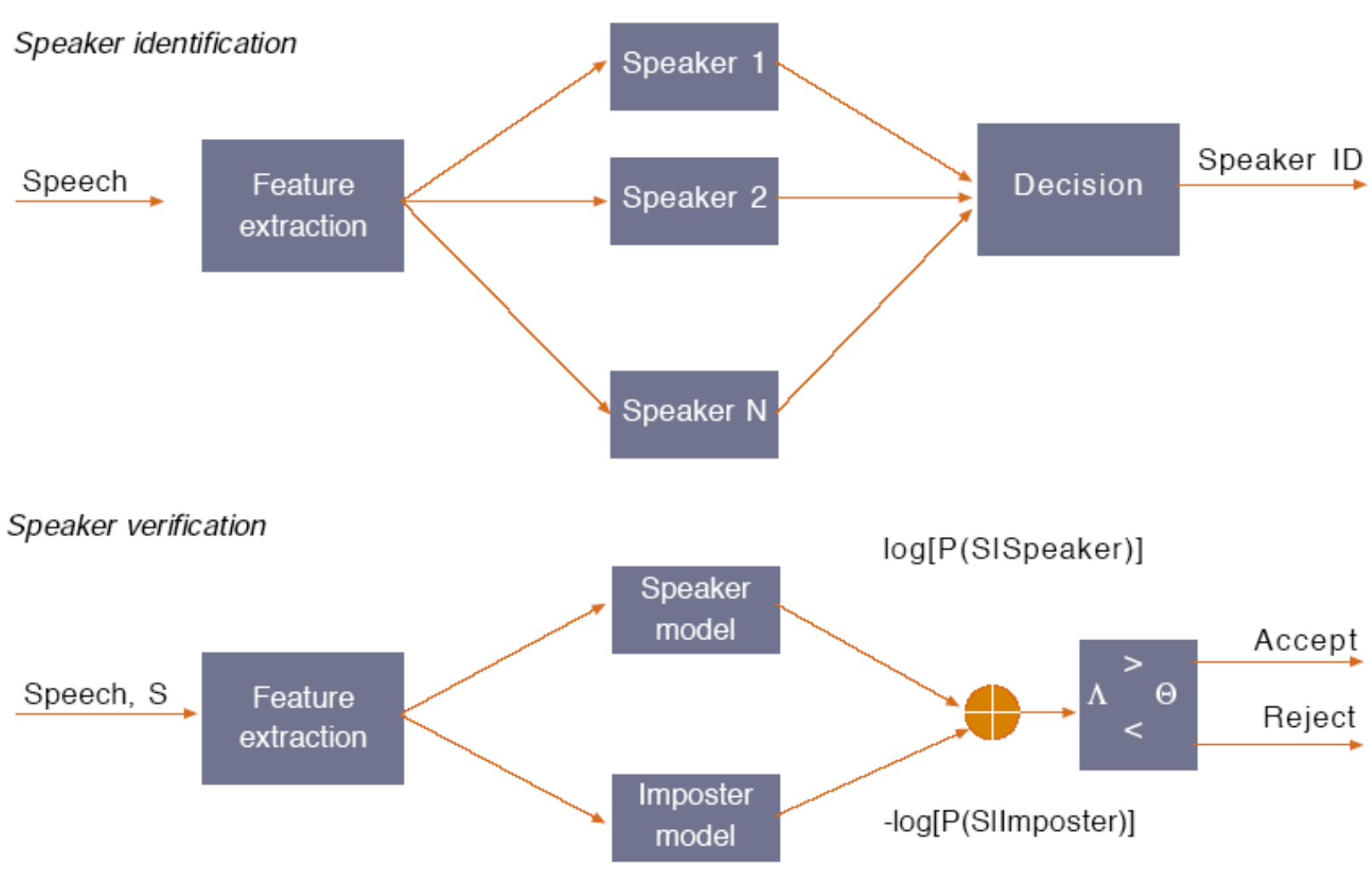}
  \end{tabular}
  \caption{Major components for speaker identification and speaker verification. \label{fig:diagram}}
\end{figure}

Fig. \ref{fig:diagram} provides digrams for speaker identificatio and verification. the main approaches in this area includes 1) template matching such as nearest neighbor \cite{higgins1994speaker} and vector quantization \cite{soong1987report}, 2) neural network, such as time delay neural network \cite{snyder2015time}, decision tree \cite{farrell1994speaker}, and 3) probabilistic models, such as Gaussian Mixture Model (GMM) with Universal Background Model (UBM) \cite{reynolds2000speaker}, joint factor analysis \cite{kenny2005joint}, i-vector \cite{dehak2011front, dehak2011language}, Support Vector Machine (SVM) \cite{campbell2006support}, etc.  Methods can be divided into text-dependent and text-independent, where the former achieves better performance with additional information, and the latter is more user friendly and easier to use. Reynolds \cite{reynolds2002overview} and Fauve \cite{fauve2007state} provided a good overview of some common speech recognition applications with the state-of-the-art performance.

This paper proposes a neural network framework for text-independent speaker classification and verification, using TIMIT 8K database. With optimization in feature and model training, the system achieves 100\% classification accuracy with slightly more than 1 second speech, and less than 6\% ERR in speaker verification with more than 100 impostor size, using approximately 5 seconds data.  

The following sections walk through the major pieces of this work, including feature engineering (Sec. \ref{sec:data}), design, implementation and results for speaker classification and verification systems (Sec. \ref{sec:nnsc} and Sec. \ref{sec:nnsv}). Finally, the conclusion and future work is given in Sec. \ref{sec:conclusion}. 

\section{Data Preparation and Feature Engineering}
\label{sec:data}

The following 3 subsections introduce the database used in this paper, and the process of converting raw speech into features used that used in speaker classification and verification, including a) preprocessing, and b) feature extraction, normalization and concatenation.

\subsection{Database}
\label{subsec: database}

Speech of all 326 male speakers from 8 different dialect regions in the ``train'' folder of the TIMIT corpus with 8K sampling rate is used here. Data of males from the ``test'' folder and data of females from both ``train'' and ``test'' folders are currently reserved for future development. For each speaker, there are 10 data files containing one sentence each with duration about 2.5 seconds. They are from 3 categories: ``SX'' (5 sentences), ``SI'' (3 sentences) and ``SA'' (2 sentences). Data are first sorted alphabetically by speaker name in their dialect region folders, then combined to form a list of data containing 326 speakers. They are then divided into 2 groups: first 200 speakers (group A) and remaining 126 speakers (group B). For speaker classification ``SX'' sentences in group A are used to train the text-independent Neural Network Speaker Classifier (NNSC), while the ``SA'' and ``SI'' sentences in group A are used to test. For speaker verification, since it is based on NNSC, only ``SA'' and ``SI'' sentences are used to avoid overlapping with any training data used in model training. Speakers in group A are used as in-domain speakers, and speakers in group B are used as out-of-domain speakers (imposters). 

\subsection{Preprocessing}
\label{subsec:preprocessing}

Preprocessing mainly consists of a) scaling the maximum of absolute amplitude to 1, and b) Voice Activity Detection (VAD) to eliminate the unvoiced part of speech. Experiments show both speaker classification and verification can perform significantly better if speakers are evaluated only using voiced speech, especially when the data is noisy.

An improved version of Giannakopoulos's recipe \cite{giannakopoulos2009method} with short-term energy and spectral centroid is developed for VAD. Given a short-term signal $s(n)$ with $N$ samples, the energy is:
\begin{equation}
\label{eq:ste}
E = \frac{1}{N} \sum_{n=1}^{N}|s(n)|^2 , 
\end{equation}
and given the corresponding Discrete Fourier Transform (DFT) $S(k)$ of $s(n)$ with $K$ frequency components, the spectral centroid can be formulated as:
\begin{equation}
\label{eq:spectral centroid}
C = \frac{\sum_{k=1}^{K}kS(k)}{\sum_{k=1}^{K}S(k)} .
\end{equation}
The Short-Term Energy (STE) $E$ is used to discriminate silence with environmental noise, and the Spectral Centroid (SC) $C$ can be used to remove non-environmental noise, i.e. non-speech sound, such as coughing, mouse clicking and keyboard tapping, since they normally have different SCs compared to human speech. When computing the frame-level $E$ and $C$, a $50$ ms window size and a $25$ ms hop size are used.

\begin{figure}[htb]
\begin{minipage}[b]{1\linewidth}
  \centering
  \centerline{\includegraphics[height=4.5cm, width=7cm]{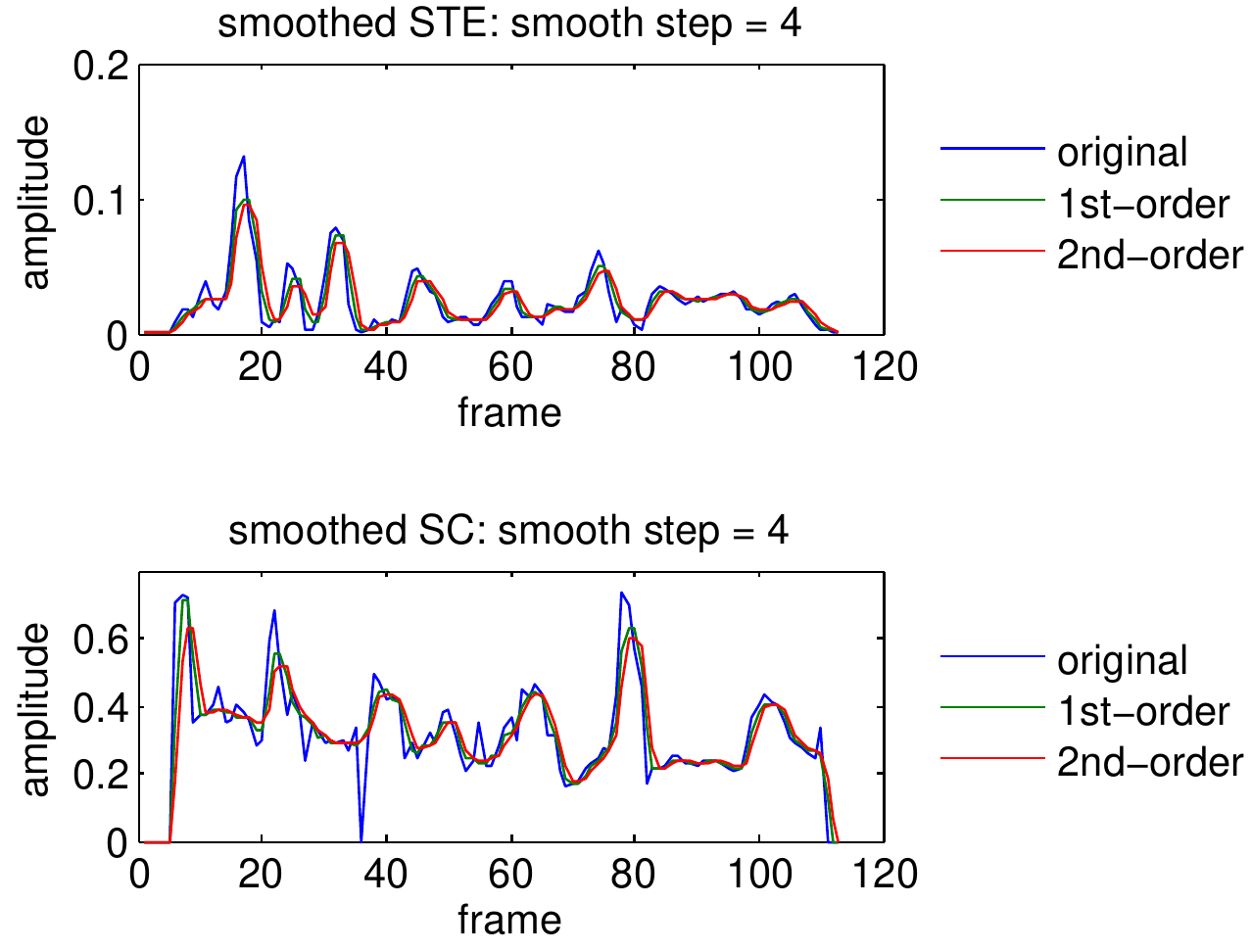}}
  \centerline{(a) smoothing step size 4}\medskip
\end{minipage}
\hfill
\begin{minipage}[b]{1\linewidth}
  \centering
  \centerline{\includegraphics[height=4.5cm, width=7cm]{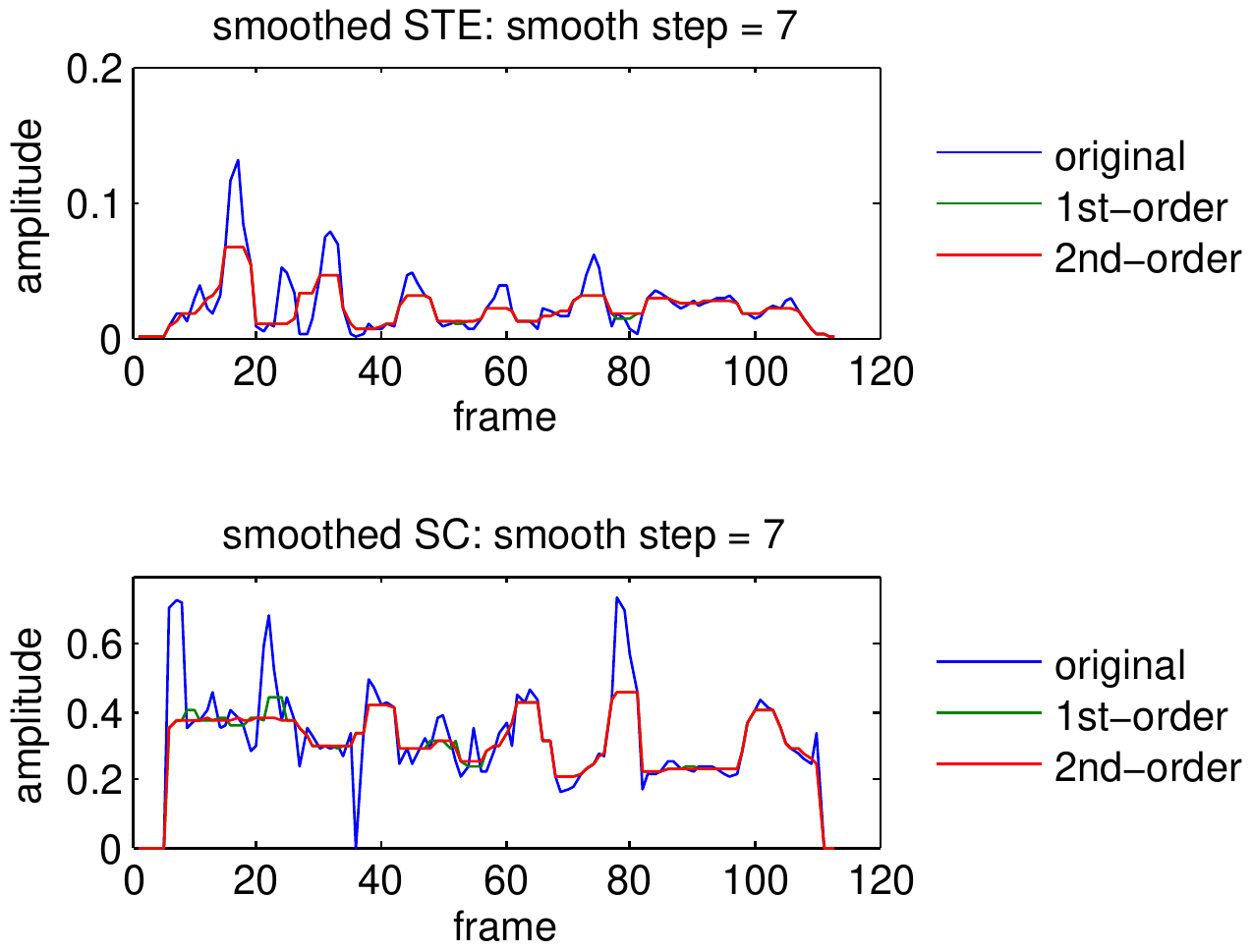}}
  \centerline{(b) smoothing step size 7}\medskip
\end{minipage}
\caption{Short-term energy and spectral centroid with different median filter smoothing steps and orders.}
\label{fig:median_step_example}
\end{figure}

To set the overall threshold, only when $E$ and $C$ are both above their thresholds $T_{E}$ and $T_{C}$, the speech frame is considered to be voiced, otherwise, it will be removed. These thresholds are adjusted to be slightly higher to enforce a stricter VAD algorithm and ensure the quality of the captured voiced sections. This is achieved by tuning the signal median smoothing parameters, such as step size and smoothing order, as well as setting the thresholds $T_{E}$ and $T_{C}$ as a weighted average of the local maxima in the distribution histograms of the short-term energy and spectral centroid respectively. Fig. \ref{fig:median_step_example} is an example of applying different median filter smoothing step sizes to STE and SC. Larger step size (e.g. 7) and order (e.g. 2) are used in order to achieve more stricter VAD. 

\subsection{Feature Extraction, Normalization and Concatenation}
\label{subsec:feature}

\begin{figure}[htb]
 \centering
  \begin{tabular}{c}
 	\includegraphics[height=6cm, width=7cm]{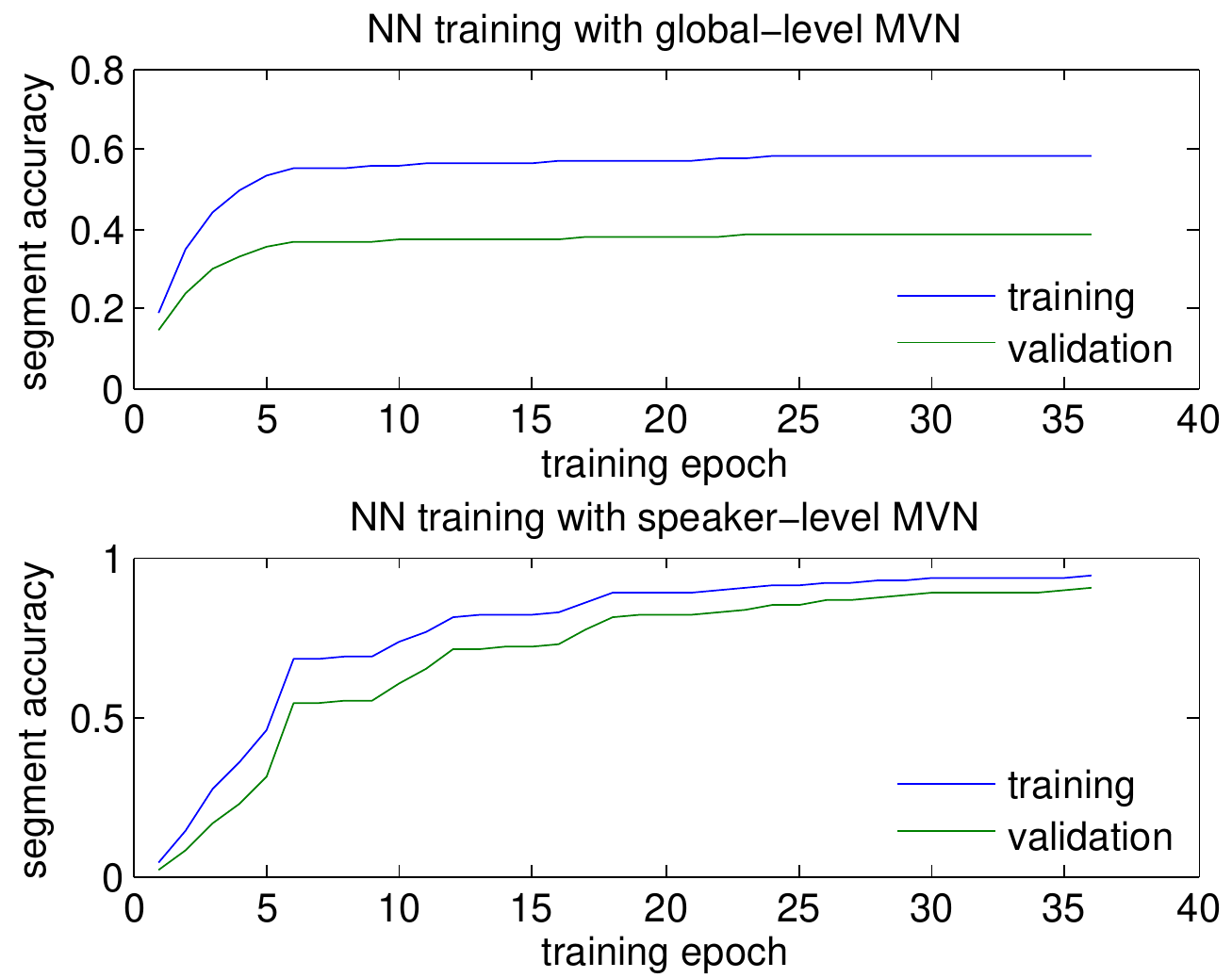}
  \end{tabular}
  \caption{Comparison of global-level MVN vs. speaker-level MVN in NN training in terms of training and validation frame accuracies. \label{fig:gmvn_vs_smvn}}
\end{figure}

The 39-dimensional Mel-Frequency Cepstral Coefficients (MFCCs) with delta and double delta were generated from the preprocessed speech, following Ellis's recipe \cite{Ellis05-rastamat}. They were extracted using overlapped $25$ ms Hamming windows which hop every $10$ ms. Then, the features of each speaker were normalized with his own mean and variance (speaker-level MVN, or SMVN), instead of using the overall mean and variance (global-level MVN, or GMVN). Fig. \ref{fig:gmvn_vs_smvn} shows SMVN though converges slower, but helps to achieve better feature frame level training and validation accuracies in network training. It is slightly counter-intuitive, since SMVN overlaps speaker patterns on top of each other. However, it can match the instances of patterns from the same speaker better than GMVN as the training goes. 

\begin{figure}[htb]
 \centering
  \begin{tabular}{c}
 	\includegraphics[height=3cm, width=6cm]{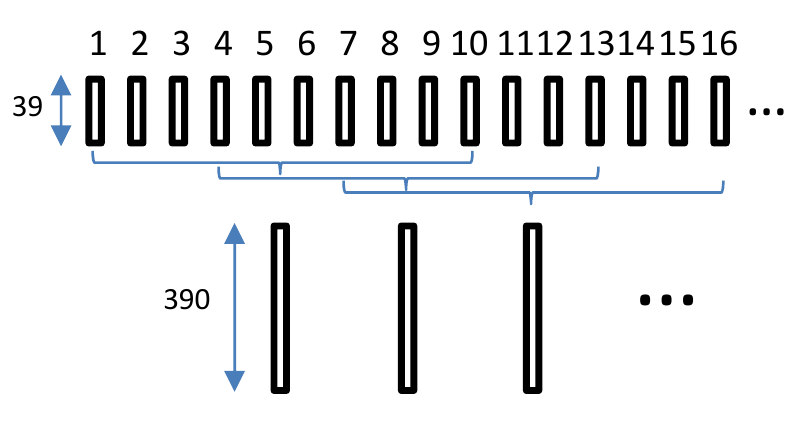}
  \end{tabular}
  \caption{Feature concatentation example with a window size of 10 frames and a hop size of 3 frames. \label{fig:feature_concatenation}}
\end{figure}

To capture the transition patterns within longer durations, these 39-dimensional feature frames were concatenated to form overlapped longer frames. In this work,  10 frames ($100$ ms) were concatenated with hop size of 3 frames ($30$ ms) as shown in Fig. \ref{fig:feature_concatenation}.   

\section{Neural Network Speaker Classification}
\label{sec:nnsc}

The concatenated features (e.g. 390 dimensional feature vectors) are used as the input to a  neural network speaker classifier. As mentioned in the first paragraph of Sec. \ref{sec:data}, the ``SX'' and ``SI'' sentences of the first 200 male speakers were used for training, and the remaining ``SA'' sentences from the same set of speakers were used for testing.

\subsection{Cost Function and Model Structures}
\label{subsec:cost}

Ng's neural network training recipe for hand-written digit classification \cite{couseraml} is used here, which treats the multi-class problem as $K$ separate binary classifications. It is considered to be the generalization of the cost function of binary classification using logistic regression, which is built on slightly different concepts compared with the cross-entropy cost function with softmax as the output layer \cite{srihari}. 

Given $M$ samples, $K$ output classes, and $L$ layers, including input, output and all hidden layers in between, the cost function can be formulated as:
\begin{eqnarray}
\label{eq:cost_function}
J(\Theta) &=& -\frac{1}{M}  \left[ \sum_{m=1}^{M}\sum_{k=1}^{K} \left( y_{k}^{(m)} \log (h_{\theta}(x^{(m)})_{k}) \right. \right. \\ \nonumber 
    & &  + \left. \left. (1-y_{k}^{(m)}) \log (1-h_{\theta}(x^{(m)})_{k}) \right) \right] \\ \nonumber
    & & + \frac{\lambda}{2M}\sum_{l=1}^{L-1}\sum_{i=1}^{s_{l}}\sum_{j=1}^{s_{l+1}}(\theta_{ji}^{(l)})^{2}
\end{eqnarray}
where $h_{\theta}(x^{(m)})_{k}$ is the $k$th output of the final layer, given $m$th input sample $x^{(m)}$, and $y_{k}^{(m)}$ is its corresponding target label. The $2$nd half of Eq.  (\ref{eq:cost_function}) is the regularization factor to prevent over-fitting, where $\lambda$ is the regularization parameter and $\theta_{ji}^{(l)}$ is the $j$-th row, $i$-th column element of the weight matrix $\Theta^{(l)}$ between $l$-th and $(l+1)$-th layers, i.e. the weight from $i$-th node in $l$-th layer to $j$-th node in $(l+1)$-th layer. 

In this work, there is only 1 hidden layer ($L=3$) with $200$ nodes ($s_{2}=200$), the input feature dimension is $390$ ($s_{1}=390$), and the speaker classifier was trained with data from $200$ speakers ($s_{3}=K=200$). Therefore, the network structure is $390:200:200$, with weight matrices $\Theta^{(1)}$ ($200 \times 391$) and $\Theta^{2}$ ($200 \times 201$). The additional 1 column is a bias vector, which is left out in regularization, since the change of bias is unrelated to over-fitting. In this example, the regularization part in Eq. (\ref{eq:cost_function}) can be instantiated as
\begin{eqnarray}
\label{eq:cost_function_example}
\sum_{l=1}^{L-1}\sum_{i=1}^{s_{l}}\sum_{j=1}^{s_{l+1}}(\theta_{ji}^{(l)})^{2} = \
\sum_{i=1}^{390}\sum_{j=1}^{200}(\theta_{j,i}^{(1)})^2 \
     + \sum_{i=1}^{200}\sum_{j=1}^{200}(\theta_{j,i}^{(2)})^2.
\end{eqnarray}
%
 

\subsection{Model Training and Performance Evaluation}
\label{subsec:model}

The neural network model is trained through forward-backward propagation. Denoting $z^{(l)}$ and $a^{(l)}$ as the input and output of the $l$-th layer, the sigmoid function
\begin{equation}
\label{eq:sigmoid}
	a^{(l)} = g(z^{(l)}) = \frac{1}{1 + e^{-z^{(l)}}}  
\end{equation}
is selected as the activation function, and the input $z^{(l+1)}$ of the $(l+1)$-th layer can be transformed from the output $a^{(l)}$ of the $l$-th layer, using $z^{(l+1)} = \Theta a^{(l)}$. Then, $h_{\theta}(x)$ can be computed through forward propagation: $x = a^{(1)} \rightarrow z^{(2)} \rightarrow a^{(2)} \rightarrow \cdots \rightarrow z^{(L)} \rightarrow a^{(L)} = h_\theta(x)$. The weight matrix $\Theta^{(l)}$ is randomly initiated using continuous uniform distribution between $(-0.1, 0.1)$ and then trained through backward propagation of $\partial{J}/\partial{\theta_{j,i}^{(l)}}$, by minimizing $J(\Theta)$ using Rasmussen's conjugate gradient algorithm, which handles step size (learning rate) automatically with slope ratio method\cite{rasmussen2006gaussian}. 

\begin{figure}[htb]
\begin{minipage}[b]{1\linewidth}
  \centering
  \centerline{\includegraphics[height=4cm, width=6cm]{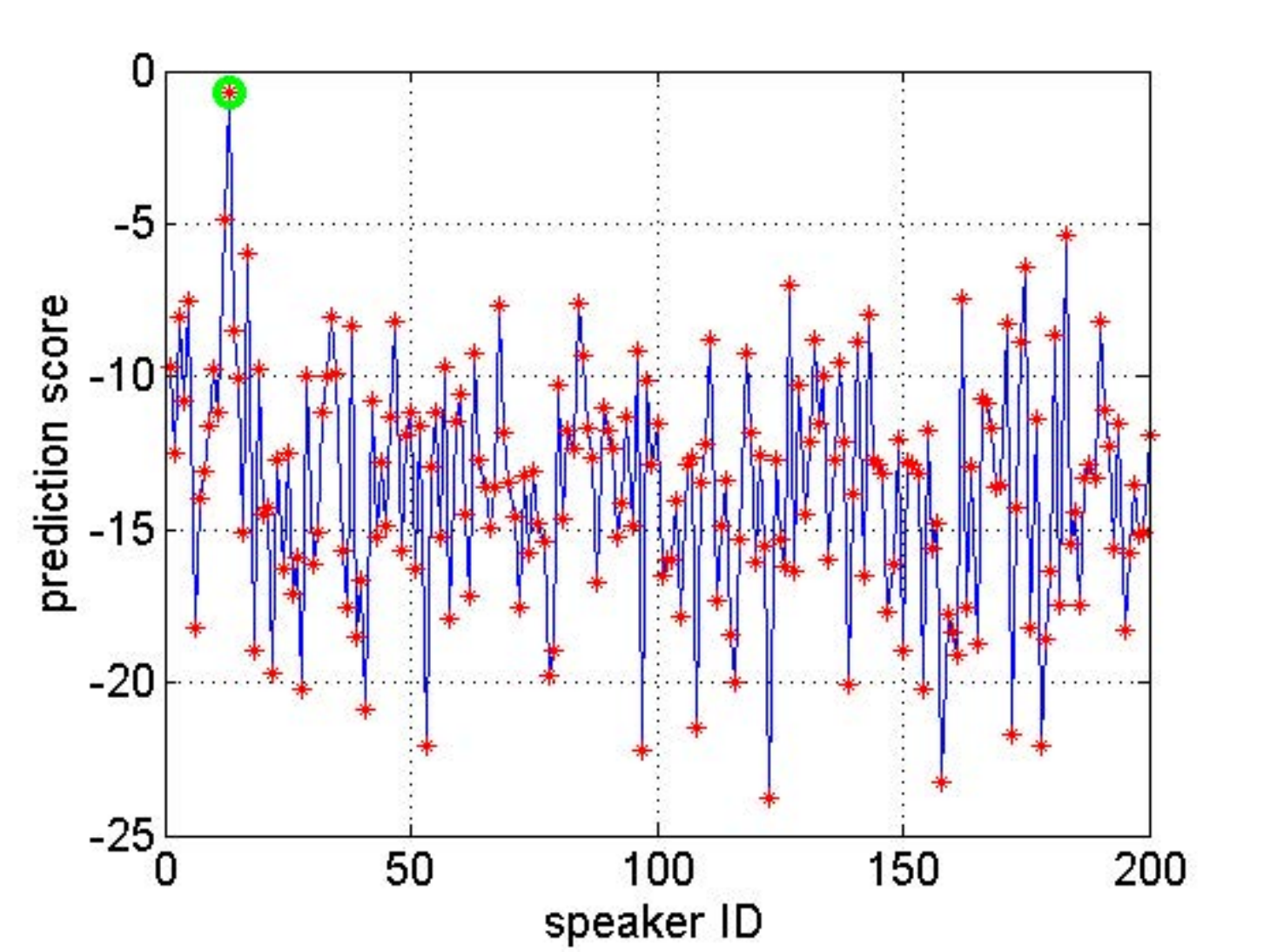}}
  \centerline{(a) {\tt SI1410} in training}\medskip
\end{minipage}
\hfill
\begin{minipage}[b]{1\linewidth}
  \centering
  \centerline{\includegraphics[height=4cm, width=6cm]{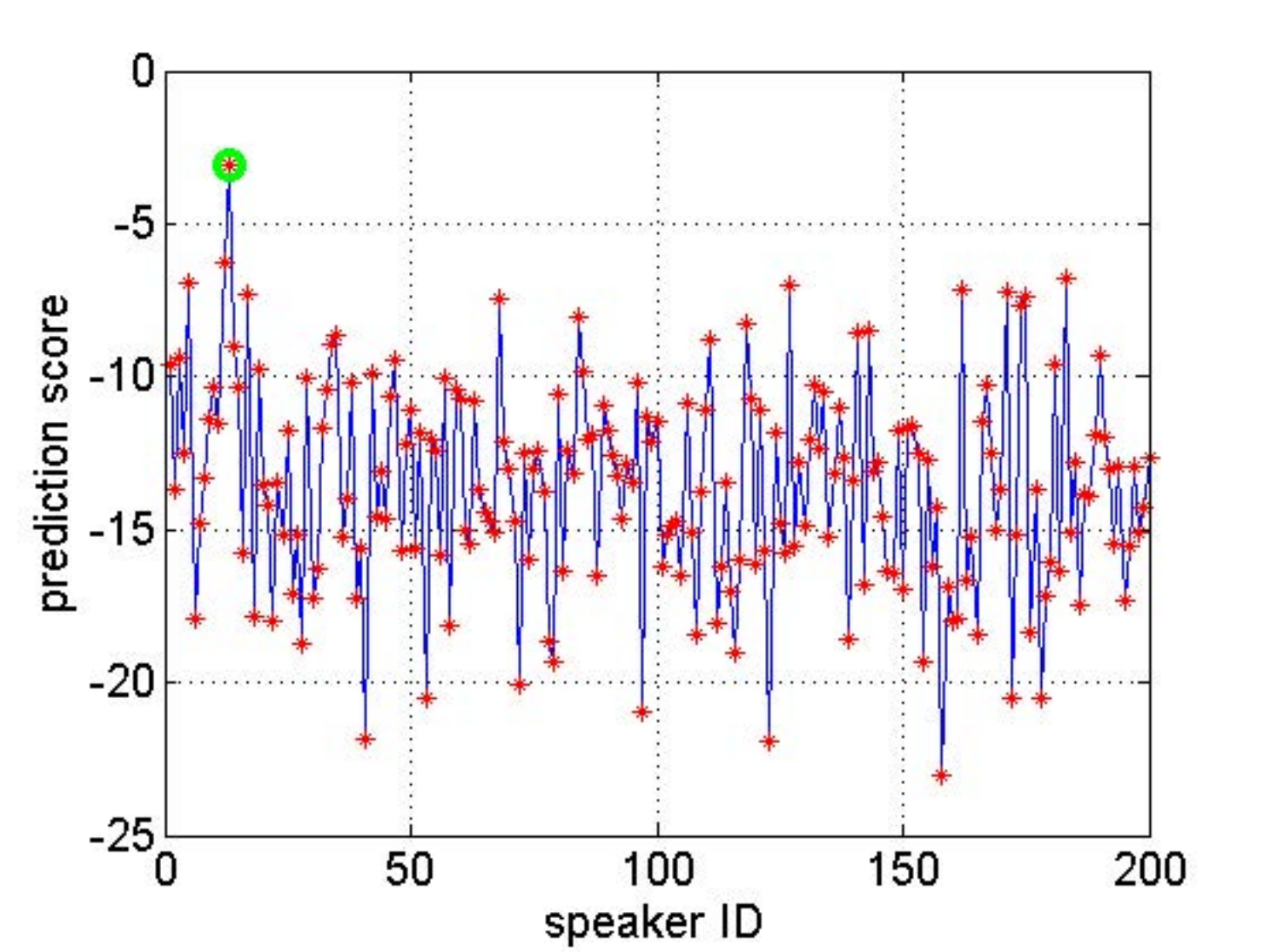}}
  \centerline{(b) {\tt SA1} in testing}\medskip
\end{minipage}
\caption{File-level prediction scores of $13$th speaker (MPGR0) in training and testing sets respectively.}
\label{fig:classifiation_example}
\end{figure}

In evaluating the classifier performance, the sigmoid output of the final layer $h_{\theta}(x^{(m)})$ is a $K$-dimensional vector, each element in the ranges of $(0,1)$. It serves as the ``likelihood'' to indicate how likely it is to classify $m$-th input frame into one of the $K$ speakers. The speaker classification can be predicted by the sum of log likelihood of $M$ input frames (prediction scores), and the predicted speaker ID $k^{*}$ is the index of its maximum:
\begin{equation}
\label{eq:prediction}
	k^{*} = \argmax_{k \in [1,K]} \left( \sum_{m=1}^{M} \log (h_{\theta}(x^{(m)})_{k}) \right) .
\end{equation}
$M$ can range from 1 to the entire frame length of the testing file. If $M=1$, the accuracy achieved is based on individual frames, each of which is $100$ ms (window duration $T_{win}$ in feature concatenation) with $30$ ms of new data, compared with the previous frame. On the other hand, if $M$ is equal to the total number of frames in file, the accuracy is file-based. The average duration of sentences (i.e. file length) is about 2.5 seconds. In general, larger $M$ leads to higher accuracy. Given the best model available with the network structure $390:200:200$, Fig. \ref{fig:classifiation_example} demonstrates an example of file-level prediction score of $13$-th speaker (MPGR0). It shows the peak of positives (in the green circle) is slightly dropped but still distinguishable enough to all other negatives, from the file {\tt SI1410} in the training set, to the file {\tt SA1} in the testing set.

Using this model, the file-level training and testing accuracies at $200$ speaker size are both 100\%, as indicated in Table \ref{tab:classification_accuracy}. 
\begin{table}[htb]
\footnotesize
  \caption{NN-based speaker classification performance with first 200 male in 8K TIMIT ($0.1$ sec./frame, $\sim$2.5 sec./file)}
  \label{tab:classification_accuracy}\centering
  \setlength{\tabcolsep}{2.25pt}
  \begin{tabular}{@{} *{6}{c} @{}} \toprule%
    \multirow{2}*{\textbf{Dataset}} &  \multicolumn{2}{c}{\textbf{Accuracy (\%)}} & \multicolumn{3}{c}{\textbf{Frame (sec.) needed for 100\% accuracy}} \\
     & \textbf{\textit{frame}} & \textbf{\textit{file}} & \textbf{\textit{min}} & \textbf{\textit{mean}} & \textbf{\textit{max}} \\ \midrule
     train & 93.29 & 100 & 2 (0.13) & 3.23 (0.17) & 5 (0.22) \\
     test & 71.42 & 100 & 6 (0.25) & 13.55 (0.48) & 37 (1.18) \\\bottomrule
  \end{tabular}
\end{table} 
The frame-level testing accuracy is $71.42$\%, which indicates that $71.42$\% frames in the testing set, with duration as little as $0.1$ second, can be classified correctly. It also shows the minimum, mean, and maximum number of consecutive feature frames needed and their corresponding durations in order to achieve 100\% accuracy, evaluated through all files in both training and testing datasets. Since the next frame provides only $30$ms (hop duration $T_{hop}$ in feature concatenation) additional information, compared with the current frame, given the number of frames needed $N$, the formula to compute the corresponding required duration $T$ is
\begin{eqnarray}
\label{eq:frame-duration}
	T = (N-1) \times T_{hop} + 1 \times T_{win} . 
\end{eqnarray}
With this formula, it requires only 13.55 frames (0.48 second) on average, to achieve 100\% accuracy in the testing dataset. 

Using the training data to test is normally not legitimate, and here it is used merely to get a sense of how the accuracy drops when switching from training data to testing data. 

\subsection{Model Parameter Optimization}
\label{subsec:optimization}

The current neural network model with the structure $390:200:200$ is actually the best one in terms of highest frame-level testing accuracy, after grid searching on a) the number of hidden layers ($1, 2$), and b) the number of nodes per hidden layer ($50, 100, 200, 400$), with a subset containing only 10\% randomly selected training and testing data. 

Once the ideal network structure is identified, the model training is conducted with a regularization parameter $\lambda$ in the cost function $J(\Theta)$, which is iteratively reduced from 3 to 0 through training. This dynamic regularization scheme is experimentally proved to avoid over-fitting and allow more iterations to reach a refined model with better performance.   


The training is set to be terminate once the testing frame accuracy cannot be improved more than $0.1\%$ in the last 2 consecutive training iterations, which normally takes around $500$ to $1000$ iterations. The training set is at $200$ speaker size with $20$ seconds speech each. It is fed in as a whole batch of data, which requires about 1 hour to train, on a computer with i7-3770 CPU and 16 GB memory. Therefore, the computational cost is certainly manageable.     

\section{Neural Network Speaker Verification}
\label{sec:nnsv}

This section first introduces the mechanism of converting speaker classification into speaker verification; then describes the method of developing speaker-specific thesholds to shift verification outputs; finally it evaluates the system with metrics such as Equal Error Rate (EER).

\subsection{Verification Mechanism}
\label{subsec:mechanism}

In speaker verification, the assumption that any input speaker will be one of the in-domain speakers is no longer kepted. When the testing speaker is claimed to be speaker $k$ and the highest output score is also from the $k$-th output nodes, he might be a imposter, who is more similar to speaker $k$, and less similar to the rest of $K-1$ enrolled (in-domain) speakers. So Eq. (\ref{eq:prediction}) in Subsec. \ref{subsec:model} is no longer hold and a threshold is necessary to determine if the testing speaker is similar enough to the targeting speaker and can be verified as speaker $k$.   

\begin{figure}[htb]
 \centering
  \begin{tabular}{c}
 	\includegraphics[height=5cm, width=6cm]{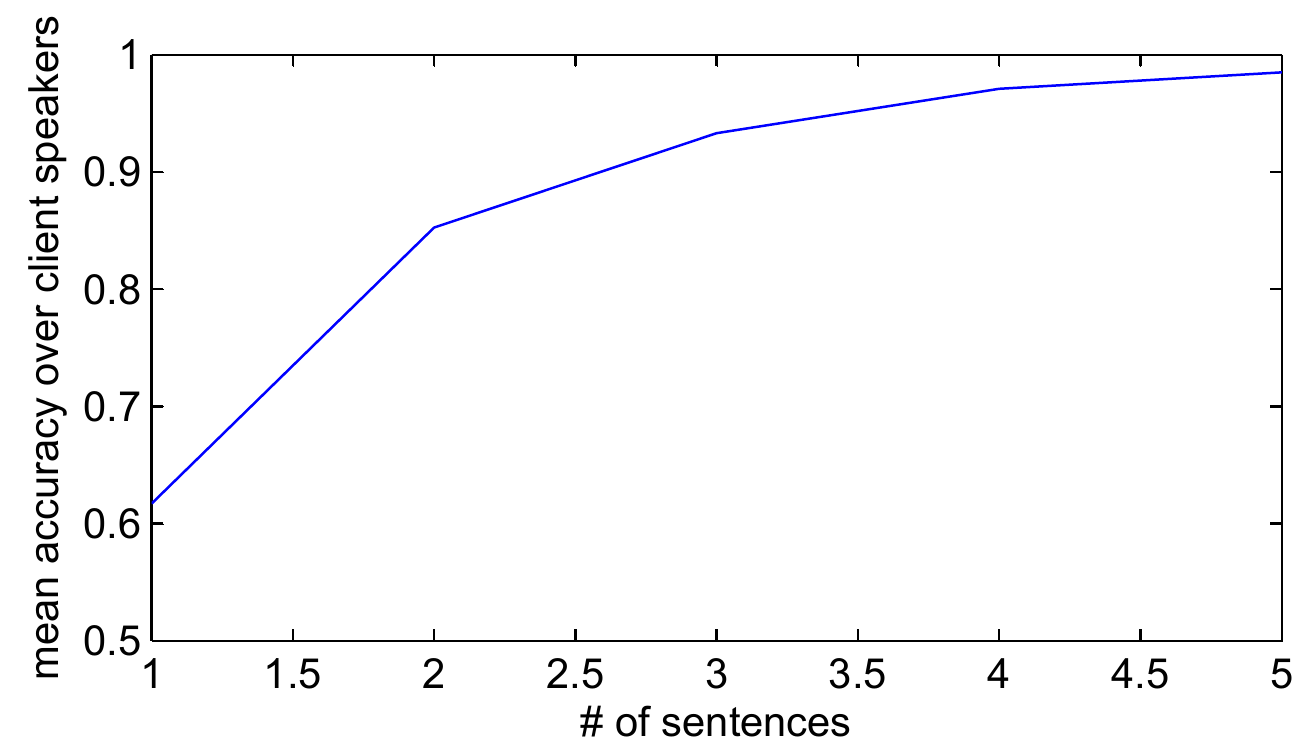}
  \end{tabular}
  \caption{Verification accuracy (1 in-domain client speaker vs. 126 out-of-domain imposters) vs. number of testing files, averaged over all 200 in-domain speakers in TIMIT. \label{fig:accuracy_vs_length}}
\end{figure}

Let the mean $K$-dimensional output prediction vector over feature frames for client speaker $k$, given features $x_{l}$ of speaker $l$ be: 
\begin{equation}
\label{eq:output}
	\textit{\textbf{O}}(k,l) = \frac{1}{M} \sum_{m=1}^{M} \log (h_{\theta}(x_{l}^{(m)})_{k}) ,
\end{equation}
where $M$ is the number of frames in the testing feature. In this project, client speakers are the first 200 male speakers in TIMIT ($K=200$), and the imposters (out-of-domain) are the ramaining 126 speakers ($L=126$). In positive verification, where $l=k$, and the $k$-th value on $\textit{\textbf{O}}(k,k)$, i.e. $\textit{\textbf{O}}_{k}(k,k)$ should be high; while in negative verification, where $l \in [1,L]$, and $\textit{\textbf{O}}_{k}(k,l)$ should be low. If
\begin{equation}
\label{eq:verification}
	\textit{\textbf{O}}_{k}(k,k) > \mathrm{any} (\textit{\textbf{O}}_{k}(k,l)), \quad l \in [1,L],
\end{equation}
then, the $k$-th speaker can be correctly verified. In our experiment, $\textit{\textbf{O}}(k,k)$ and $\textit{\textbf{O}}(k,l)$ are actually normalized over $K$ output node dimension, and the normalized versions are:
\begin{equation}
\label{eq:normalization}
	\textit{\textbf{O}}^{'}(k,k) = \frac{\textit{\textbf{O}}(k,k)}{\sum_{k=1}^{K}\textit{\textbf{O}}(k,k)} ,  \textit{\textbf{O}}^{'}(k,l) = \frac{\textit{\textbf{O}}(k,l)}{\sum_{k=1}^{K}\textit{\textbf{O}}(k,l)} .
\end{equation}
It is found to achieve better verification accuracy by penalizing the ones with strong competing speakers. Fig. \ref{fig:accuracy_vs_length} shows the accuracy vs. number of testing files (up to 5 since there are 5 sentences from ``SI'' and ``SA'' categories). For example, the mean accuracy is $61.7\%$ when speakers are tested with individual files and $85.25\%$ when tested with a combination of two files ($\binom 52=10$ cases). The sentences duration is about 2.5 seconds each, so it is similar to the accuracy with testing duration 2.5 seconds, 5 seconds, etc. For each out of the 200 client speakers, the accuracy is binary, either 1, i.e. Eq. (\ref{eq:verification}) is satisfied, or 0 otherwise.

\subsection{Speaker Specific Thresholding}
\label{subsec:threshold}

The accuracy measurement above will drop significantly when the imposter size is getting larger. In fact, it is merely an analysis to demonstrate the challenge to maintain high accuracy with a large imposter size which is rare in the real scenario. Next, the speaker-specific thresholds will be obtained by finding the Gaussian distributions of the positive (testing speaker is the client speaker) and negative (testing speakers is one of the imposters) samples, using Bayes rule.

\begin{figure}[htb]
 \centering
  \begin{tabular}{c}
 	\includegraphics[height=5cm, width=6cm]{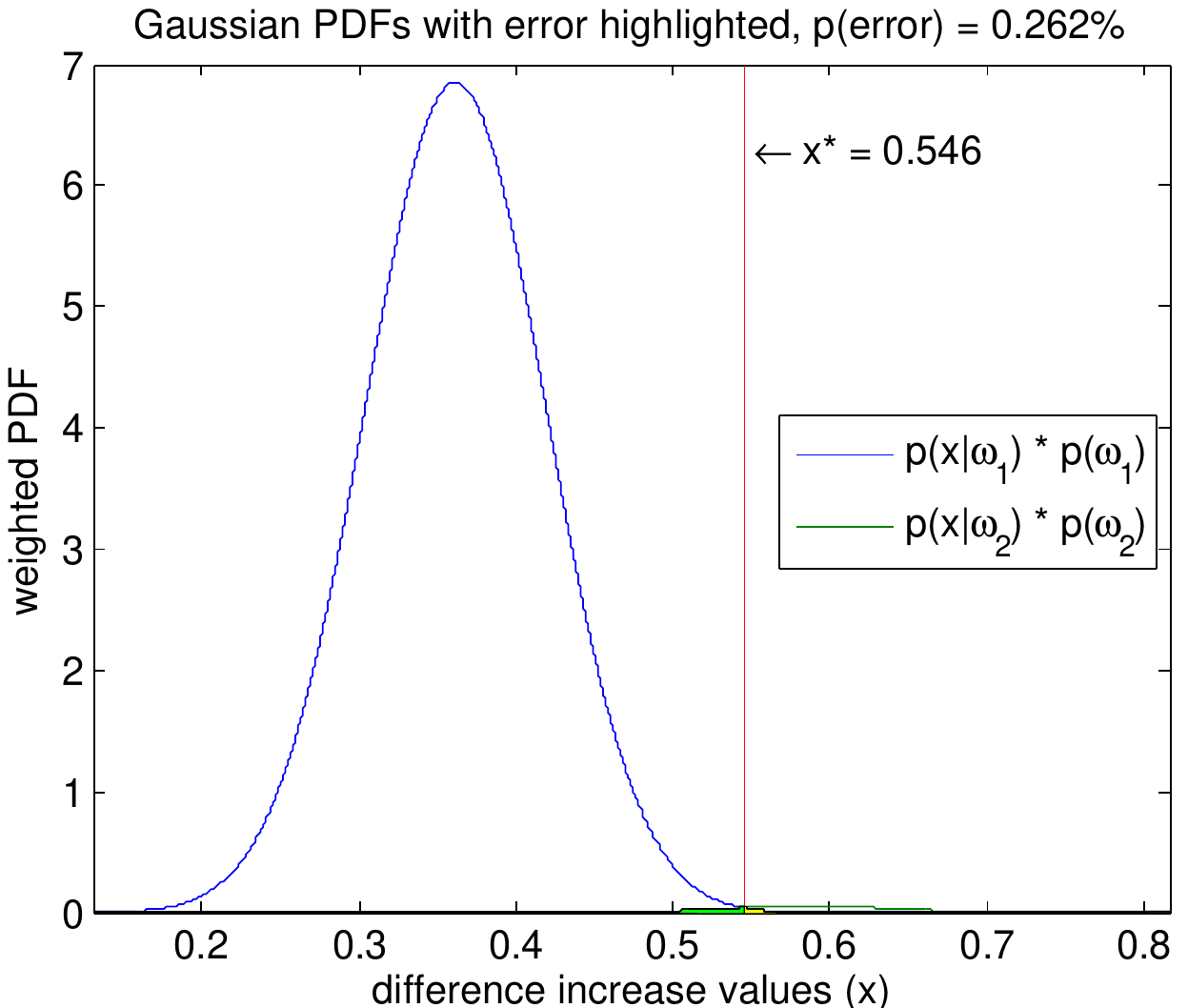}
  \end{tabular}
  \caption{Example of thresholding with 2 Gaussians distributions of positive and negative samples. Sample values are collected with combinations of 2 files (10 cases with $\sim5$ seconds in duration), i.e. 10 positives vs. 1260 negatives. \label{fig:thresholding}}
\end{figure}
    
Since the positive and negative is extremely skewed with current 126 imposter size (i.e. positive:negative is 1:126), the distribution for the positive samples has a very low prior and almost invisible in Fig. \ref{fig:thresholding}. However, the estimated threshold, which is the intersection of the two Gaussians, can be still found by solving the Eq. (\ref{eq:intersection}) using the root finding method, which first reformats the Eq. (\ref{eq:intersection}) to quadratic function $ax^2 + bx + c = 0$, and then represents $x$ by $a,b,c$.
\begin{equation}
\label{eq:intersection}
	\frac{p_{1}}{\sigma_{1}} e^{\frac{(x-u_{1}^2)}{2\sigma_{1}}} = \frac{1-p_{1}}{\sigma_{2}} e^{\frac{(x-u_{2})^2}{2\sigma_{2}}} .
\end{equation} 

\subsection{Performance with Optimized Thresholds}
\label{subsec:performance}

\begin{figure}[htb]
 \centering
  \begin{tabular}{c}
 	\includegraphics[height=5cm, width=6cm]{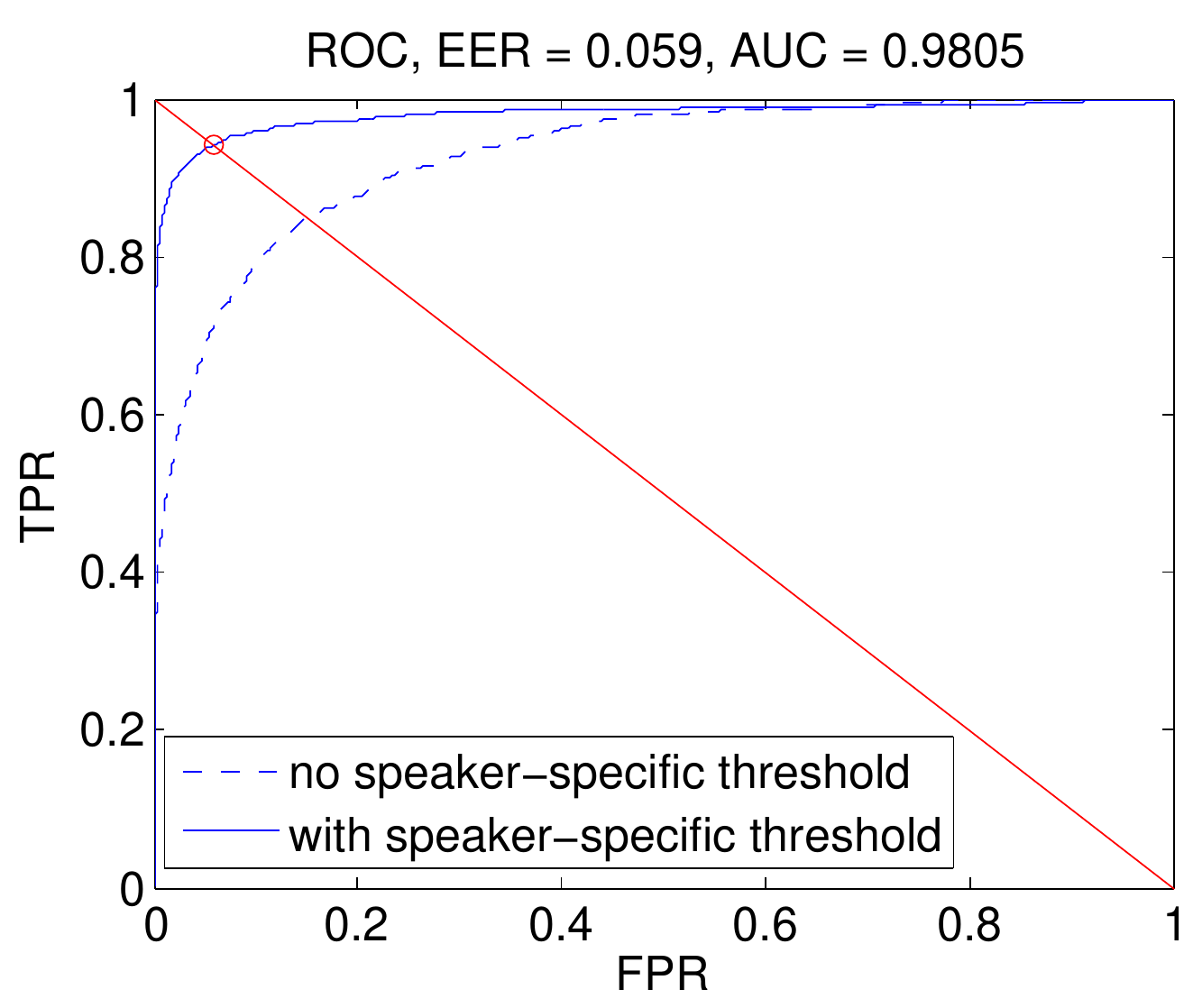}
  \end{tabular}
  \caption{ROC when verifying with length of 2 files ($\sim5$ seconds), with or without speaker-specific thresholds. \label{fig:roc}}
\end{figure}

With the speaker-specific thresholds $T_{k}$, $k \in [1,K]$, the output normalized prediction vector is shifted by 
\begin{equation}
\label{eq:offset}
	\textit{\textbf{O}}^{'}(k,l) \rightarrow \textit{\textbf{O}}^{'}(k,l) - T_{k}, l \in \{k, [1,L]\}.
\end{equation} 
Then, ROC curve is computed to find the Equal Error Rate (EER), which is a common performance indicator to evaluate biometric systems. EER equal to False Positive Rate (FPR), when $FPR + TPR = 1$. Fig. \ref{fig:roc} demonstrates the ROC curve, when verifying with length of 2 files ($\sim5$ seconds). By offsetting outputs with speaker-specific thresholds, the EER is reduced from $14.9\%$ to $5.9\%$. Another metric Area Under Curve (AUC) is $98.05\%$, and the global threshold corresponding to this best EER is $-0.0941$. 

\section{Conclusion and Future Work}
\label{sec:conclusion}

This work demonstrated a novel neural net framework for speaker classification and verification with enhanced features. The performance is tested using TIMIT corpus with 8K sampling rate. For speaker classification, 200 speakers can be classified correctly with data no more than 1.18 seconds; For speaker verification, the EER is 5.9\%, when verifying 200 in-domain speakers with 126 imposters, using speech about 5 seconds long (2 TIMIT files). Though the performance of speaker classification and verification systems is difficult to compare, due to various database condition, and enrollment and testing scenarios \cite{reynolds2002overview}, 100\% classification rate using about 1 second audio and less than 6\% EER using 5 seconds data in speaker verification, is still among one of the very competitive performances in most of the cases \cite{fauve2007state}.

This is achieved by combining all the essential components, including 1) feature engineering, such as VAD/silence removal, speaker-level MVN, feature concatenation to capture transitional information, etc., 2) neural network setup, model parameter optimization, training with dynamically reduced regularization parameter in speaker classification, and 3) output score normalization and speaker-specific thresholding in speaker verification.

There is still much room for potential improvement. First, the enrollment process is typically one-by-one, rather than enrolling a group of speakers as a whole, so the recursively model training and updating need to be addressed. Second, more challenging and noisy database should be considered to added in, in order to deal with channel normalization and system robustness. Third, combining current neural network approaches with other state-of-the-art methods, such as GMM-UBM \cite{reynolds2000speaker} and i-vector \cite{dehak2011front, dehak2011language} is also desired.

\bibliographystyle{IEEEbib}
\bibliography{references}

\end{document}